\documentstyle[12pt]{article}
\begin{document}
\renewcommand{\Re}{{\rm Re \,}}
\renewcommand{\Im}{{\rm Im \,}}
\newcommand{\Tr}{{\rm Tr \,}}
\newcommand{\beq}{\begin{equation}}
\newcommand{\fre}{F[\vec{e},\vec{m}]}
\newcommand{\eeq}{\end{equation}}
\newcommand{\bea}{\begin{eqnarray}}
\newcommand{\eea}{\end{eqnarray}}
\newcommand{\dirac}{/\!\!\!\partial}
\newcommand{\Dirac}{/\!\!\!\!D}
\begin{titlepage}
\begin{center}
\hfill hep-th/9406128 \vskip .01in \hfill NYU-TH-94/06/02, IFUM/472/FT
\vskip .01in \hfill SISSA 80/94/EP, RI-6-94
\vskip .4in
{\large\bf Non-Abelian Strong-Weak Coupling Duality in (String-Derived)
N=4 Supersymmetric Yang-Mills Theories}
\end{center}
\vskip .4in
\begin{center}
{\large Luciano Girardello}$^\clubsuit$,
{\large Amit Giveon}$^\diamondsuit$,
{\large Massimo Porrati}$^\heartsuit$, and
{\large Alberto Zaffaroni}$^\spadesuit$
\vskip .1in
$\clubsuit$ Dipartimento di Fisica, Universit\`a di Milano, via Celoria 16,
20133 Milano, Italy
\vskip .05in
$\diamondsuit$ Racah Institute of Physics, The Hebrew University, Jerusalem
91904, Israel
\vskip .05in
$\heartsuit$ Department of Physics, NYU, 4 Washington Pl.,
New York, NY 10003, USA
\vskip .05in
$\spadesuit$ SISSA - International School for Advanced Studies, Via Beirut 2,
34100 Trieste, Italy
\end{center}
\vskip .4in
\begin{center} {\bf ABSTRACT} \end{center}
\begin{quotation}
\noindent
We study strong-weak coupling duality (S-duality)
in N=4 supersymmetric non-Abelian Yang-Mills theories. These theories arise
naturally as the low-energy limit of four-dimensional toroidal
compactifications of the heterotic string.

Firstly, we define the free energy in the presence of electric and magnetic
fluxes using 't Hooft's prescription, i.e. through functional integrals at
finite volume in the presence of twisted boundary conditions. Then, we find
those free energies in two limiting cases: small and large coupling
constant.

Finally, we extend the free energies to all values of the coupling constant
(and the theta angle) by presenting a fully S-duality invariant ansatz.
This ansatz obeys all relevant consistency conditions; in particular, it obeys
't Hooft duality equations and Witten's magnetic-electric
transmutation.

The existence of an S-duality invariant,  consistent definition of free
energies supports the claim made in the literature that S-duality is a
duality symmetry of N=4 SUSY Yang-Mills.
Our ansatz also suggests that N=4, irrespective of
whether partially broken or not, is in a self-dual phase: no phase
transitions occur between the strong and weak coupling regimes.
\end{quotation}
\vfill
\end{titlepage}
\noindent
String theories compactified on a circle are known to possess
duality invariances: their spectrum and interactions are invariant under
a transformation that maps the radius of the circle, $R$, into $\alpha'/R$
($\alpha'$ is the string tension). This symmetry holds at {\em each order}
in the string loop expansion, and, in general, interchanges light states
with ultra-massive ones. It is therefore essentially non perturbative in
$\alpha'$ and it is sometimes called target-space duality.

Recently another type of duality symmetry has been conjectured. It is
called strong-weak coupling duality, or S-duality~\cite{qu}.
It is a symmetry of
four-dimensional strings, which acts by transforming the four-dimensional
dilaton field $S$ as
\beq
i S \rightarrow \frac{a i S + b}{c i S +d}, \;\;\; \left( \begin{array}{cc}
 a & b \\ c & d \end{array} \right) \in {\rm SL(2,}Z{\rm )}.
\label{1}
\eeq
Since the real part of $S$ multiplies the gauge kinetic term in the 4-D
effective action of strings, this symmetry contains a duality
$S \rightarrow 1/S$, which transforms, for instance, a
weakly coupled theory ($\Re S \equiv 4\pi/g^2 \gg 1$) into a strongly
coupled one. Thus S-duality, unlike the previously mentioned $R\rightarrow
\alpha'/R$ duality, is essentially non-perturbative in the gauge coupling
constant: it cannot hold order by order in the loop expansion.

The most important phenomenological applications of S-duality regard N=1 4-D
string vacua; on the other hand, the best chance to {\em prove} S-duality
lies in toroidal compactifications of the heterotic string. These
compactifications possess an unbroken N=4 supersymmetry. Arguments in favor
of S-duality have been given almost exclusively for N=4 compactifications.
They fall into two groups. One of them conjectures a map between strings and
five-branes~\cite{5b};
under this map the role of S-duality and the target-space duality
are reversed, in particular, S-duality holds order by order in the
five-brane perturbation expansion.
The other deals with the low-energy
effective theory of the heterotic string,  obtained in the limit
$\alpha'\rightarrow 0$, $g$= constant (the flat limit).
If S-duality exists, it must holds order by order in the $\alpha'$
expansion,
in particular it
must be a symmetry of the flat limit.
Thus the study of this limit provides us with {\em necessary}
conditions for the existence of S-duality. The flat limit of a heterotic
N=4 compactification is a N=4 supersymmetric gauge theory.
Arguments exist about S-duality
of this flat limit when the gauge group (U(1)$^6\times$ SO(44)) is fully
broken to U(1)$^{28}$. In this latter, Abelian case S-duality is closely
related to the charge-monopole duality~\cite{mo}.

In this paper, we would like to
bring other arguments in favor of S-duality, always in the flat limit
$\alpha' \rightarrow 0$, by proposing a way to study
this symmetry even when the unbroken gauge group is non-Abelian.
We think that this study of non-Abelian S-duality is important {\em per se},
even without reference to strings, in that it may shed light on the wider
problem of finding the dynamical
realization of gauge symmetry in the case of gauge theories with extended
supersymmetry and, in particular, of finite theories.

A N=4 super Yang-Mills theory is completely determined by the gauge group.
The Lagrangian fields form a supermultiplet $(A_\mu^a, \lambda^a_I,
\phi^a_{IJ})$ made of  1 spin-one, 4 spin-1/2, and 6
spin-zero fields,
all in the adjoint of a gauge group $G$. The index $a$ labels the adjoint
representation, whereas the indices $I,J=1,2,3,4$ are the so-called
extension indices; finally the $\lambda^a_I$ are Weyl spinors and the
scalars obey a reality condition $2\phi_{IJ}^a=
\epsilon_{IJKL}(\phi_{KL}^a)^*$.
The non-constant
fluctuations of the dilaton $S$ are $O(\alpha'^{1/2})$ since its
kinetic term is $O(\alpha'^{-1})$. Thus, in the flat limit,
$S$ becomes indistinguishable from a bona-fide coupling constant.
The Lagrangian reads
\bea
L &=& \frac{1}{16\pi}\Re S [F_{\mu\nu}^aF^{a\,\mu\nu}
+ \bar{\lambda}^{a\, I}\Dirac
\lambda_I^a + f_{abc}\bar{\lambda}^{a\, I}\phi_{IJ}^b\lambda^{c\, J}
+ D_{\mu}\phi^{a\, IJ}D^{\mu}\phi^a_{IJ}
\nonumber \\ & &  +
  f_{abc}f_{ade}\phi^b_{IJ}\phi^{c\,JK} \phi^d_{KL}\phi^{e\,LI}] +
\frac{1}{16\pi}\Im S F_{\mu\nu}^a\tilde{F}^{a\,\mu\nu}.
\label{2}
\eea
Here $\phi^{a\, IJ}\equiv (\phi^a_{IJ})^*$, and $f_{abc}$ are the structure
constants of $G$. The Cartan-Killing metric is $\delta_{ab}$.
By Eq.~(\ref{2}), one reads off the relation between $S$, the coupling
constant
$g$, and theta angle $\theta$:
\beq
\Re S = \frac{4\pi}{g^2},\;\;\; \Im S=\frac{\theta}{2\pi}.
\label{2a}
\eeq
Thus, S-duality implies, among other things, that N=4 super Yang-Mills is
invariant under $g\rightarrow 4\pi/g$. It may seem surprising that a
non-Abelian gauge theory may possess a strong-weak coupling symmetry. This
is less surprising if one considers the case in which the non-Abelian
symmetry is fully broken to its Cartan subgroup U(1)$^{r}$, ($r$= rank $G$).
This is achieved by giving a generic VEV to the scalars present in the
theory, along one of the flat directions of the scalar potential in
Eq.~(\ref{2})\footnotemark.
\footnotetext{It is important to notice that in N=4  these directions remain
flat to all orders of perturbation theory, and, most probably, also
non-perturbatively~\cite{svv}.}
This broken gauge theory possesses stable, finite-energy monopoles of
magnetic charge $4\pi/g$~\cite{ps,o,r}.
Their mass saturates a bound due solely to the supersymmetry algebra,
and therefore receive no corrections in perturbation theory~\cite{wo}.
Thus one may expect the theory to be invariant under the duality
$S\rightarrow 1/S$, at least for $\Im S=0$, and
when accompanied by the interchange of electric
charges with magnetic monopoles. A wealth of other arguments
supporting full S-duality were given in refs.~\cite{s,ss}.
In those references, the fact that the unbroken gauge group is Abelian
allows for a unambiguous definition of the electric and magnetic fluxes (and
charges). The problem of studying N=4 theories when the unbroken gauge group
is non-Abelian remains open. In equivalent terms, within the formalism of
refs.~\cite{s,ss}, it is not clear how to study a theory at length scales
$L\ll 1/M$, where $M$ is the scale of gauge symmetry breaking.
Besides, another problem has to be addressed. As mentioned before, in
arguing for SL(2,$Z$) S-duality one makes use of the {\em Abelian}
definition of electric and magnetic charges. This definition is singular
when the gauge group is unbroken~\cite{th1,ps}. A non-singular, gauge invariant
definition of the fluxes has been given by 't Hooft in~\cite{th2}.
The 't Hooft proposal applies to any gauge theory which contains only
elementary
fields in the adjoint representation of the gauge group and, in particular, to
N=4. Let us recall the main features of this proposal.
Here we shall study the simplest non-Abelian gauge group, SU(2).\footnotemark
\footnotetext{The study
of general Lie groups (in particular SO(44), the maximal gauge group of 4-D
toroidal compactifications of the heterotic string) is conceptually identical
to the SU(2) case, but fraught with technical complications, and will be
reported in a forthcoming publication~\cite{ggpz}.}

One begins by evaluating the Euclidean functional integral in a box of sides
$(a_1,a_2,a_3,a_4)$.
\beq
W[\vec{k},\vec{m}]=\int [dA^a_\mu d\lambda^a_I d\phi_{IJ}^a] \exp (-\int d^4x
L), \;\;\; k_i\equiv n_{4i}, \;\;\; n_{ij}\equiv \epsilon_{ijk}m_k,
\label{3}
\eeq
where $i,j,.. =1,2,3$ label the spatial coordinates.
The Lagrangian $L$ is given in Eq.~(\ref{2}). The boundary conditions for
all fields are periodic up to a gauge transformation
\beq
\Phi (x + a_\nu e_\nu)=\Phi^{\Omega_\nu(x)}(x),
\label{4}
\eeq
where $e_\nu $ denotes a canonical versor of $R^4$ and repeated indices
are not summed. $\Phi$ and
$\Phi^{\Omega}$
denote generically a field of the supermultiplet and its gauge transform
under $\Omega$, respectively.
The integers $n_{\mu\nu}$ in Eq.~(\ref{3}) are defined modulo two by the
consistency condition~\cite{th2}
\beq
\Omega_\mu(x + a_\nu e_\nu)=(-1)^{n_{\mu\nu}}\Omega_\mu (x),\;\;\;
n_{\mu\nu}=-n_{\nu\mu}.
\label{5}
\eeq
$(-1)$ is the generator of the center of SU(2).
't Hooft has shown that the twist in the space directions $n_{ij}$
corresponds to a magnetic flux $2 m_i=\epsilon_{ijk}n_{jk}$.
The twist $n_{i4}$ is linked to the electric flux by the the equation
\beq
\exp\{-\beta F[\vec{e},\vec{m}]\}= \frac{1}{8} \sum_{\vec{k}\in (Z_2)^3}
\exp(\pi i \vec{e}\cdot \vec{k}) W[\vec{k},\vec{m}].
\label{6}
\eeq
Here $a_4 \equiv \beta$ and $F[\vec{e},\vec{m}]$ is the free energy of a
configuration with electric flux $\vec{e}=(e_1,e_2,e_3)$ and magnetic flux
$\vec{m}$. Notice that the fluxes are defined modulo two, that is, modulo
the order of the center ($Z_2$) of SU(2).

The behavior of these free energies in the thermodynamical
limit $a_i\rightarrow \infty$ determines the phases of a gauge theory.
The $F[\vec{e},\vec{m}]$ can be computed also in the Hamiltonian formalism,
for instance, in the gauge $A_0=0$~\cite{th2}. They read
\beq
e^{-\beta\fre}= \Tr_{\vec{m}} {\cal P
}[\vec{e}\,]
 e^{-\beta H},\;\;\; {\cal P}[\vec{e}\,]=\frac{1}{8} \sum_{\vec{k}\in (Z_2)^3}
e^{-\pi i \vec{e}\cdot\vec{k}} \prod_{i=1}^3 T_i^{k_i}.
\label{8}
\eeq
The symbol $\Tr_{\vec{m}}$ denotes the trace over gauge fields obeying
boundary conditions twisted by $\vec{m}$. ${\cal P}[\vec{e}\,]$ is the
projector onto states with definite value of the electric flux; this
projector can be
generated by the gauge transformations $T_i$, which are
periodic up to the nontrivial element of the center of SU(2).
The $T_i$ act non-trivially on
physical states, but trivially on {\em local} physical states. They obey the
following boundary conditions:
\beq
T_i(\vec{x} + \vec{v})= (-1)^{\delta_{ij}}T_i(\vec{x}),\;\;\;
v_k=\delta_{kj}a_j, \;\;\; {\rm no\; sum \; on }\; j.
\label{7}
\eeq

In general, it is impossible to evaluate the free energies in a closed form.
This computation becomes possible in two limiting cases: \\
a) $\vec{m}=0$, $g\ll 1$, and $\theta=0$;\\
b) $\vec{e}=0$, $g \gg 1$, and $\theta=0$.

Let us compute at first case a). Since the coupling constant is small, we
can compute $F[\vec{e},0]$ by using perturbation theory. This possibility
exists since N=4 super Yang-Mills has vanishing beta function~\cite{ro,m,svv}.
Thus, the renormalized coupling constant does not depend on the size of the
box. Therefore a theory with $g\ll 1$ is weakly interacting at {\em any}
length scale\footnotemark.
\footnotetext{This argument shows that a perturbative calculation of
$F[\vec{e},0]$ is unreliable in any asymptotically free theory, as for
example pure Yang-Mills.}
We had to set $\vec{m}$ to zero since in this regime
the magnetic flux
is expected to interact strongly, with coupling constant $O(1/g)\gg 1$.
At $\vec{m}=0$ all fields in the box obey periodic boundary conditions, up
to a periodic, globally defined gauge transformation; therefore, in a given
gauge they may be
expanded in Fourier series
\beq
\Phi(\vec{x})= \frac{1}{V}\sum_{p} e^{i \vec{p}\cdot \vec{x}}
\Phi_{\vec{p}},\;\;\; p^i=\frac{2\pi}{a^i}n^i,\; n^i\in Z,\;\;\; V\equiv
a_1a_2a_3.
\label{9}
\eeq
In the box, all non-zero Fourier modes have energies $O(V^{-1/3})$, moreover,
the potential energy scales as $1/g^2$. For instance, the gauge fields
possess a ``magnetic'' energy $\int d^3 x g^{-2}\vec{B}^2$.
The zero modes belonging to the
Cartan subalgebra of SU(2), instead, have energies $O(g^2V^{-1/3})$.
At $g\ll 1$ only this latter set gives a significant contribution to the
free energy~\cite{l,w}. The zero modes belonging to the Cartan subalgebra of
SU(2) read, in a general gauge ($\Phi\equiv \sum_a \Phi_a \tau_a$,
$\tau_a=\sigma_a/2$):
\bea
\vec{A}(\vec{x})&=&\Omega(\vec{x})^\dagger [\nabla + \tau_3\vec{c}\,]
\Omega(\vec{x}),\;\;\Omega(\vec{x})=e^{i\tau_a\omega_a(\vec{x})}, \nonumber \\
\lambda_{\alpha}^I(\vec{x})&=&\Omega(\vec{x})^\dagger \tau_3 a_{\alpha}^I
\Omega(\vec{x}),\;\;\;
[a^{I\,\dagger}_\alpha ,a^J_\alpha ]_+=\delta^{IJ}\delta_{\alpha\beta},
\;\;\; [a_\alpha^I , a^J_\beta ]_+=0, \nonumber \\
\phi^{IJ}(\vec{x})&=& \Omega(\vec{x})^\dagger
\tau_3\phi^{IJ}\Omega(\vec{x}).
\label{11}
\eea
Here $\alpha,\beta=1,2$ are Weyl indices.
Notice that the constant gauge configurations $\vec{c}$
(or torons~\cite{l}) are
defined only up to periodic gauge transformations. Thus, in a box of sides
$\vec{a}$, they parametrize a
compact space obtained by the identifications
\beq
 c_i \approx c_i + \frac{4\pi}{a_i},\;\;\; c_i \approx -c_i.
\label{11a}
\eeq
The first identification is due to the gauge transformation
$\Omega(\vec{x})=\exp (4\pi i \tau_3 x_i/a_i)$, the second is brought about
by the gauge transformation $\Omega=2\tau_2$,
generating G-parity (which is the Weyl group of SU(2)).
The Lagrangian of N=4 super Yang-Mills reduced to the zero modes reads, in
our normalizations
\beq
L={V\over 2 g^2}\int dt \left[ \left(\frac{d\vec{c}}{dt}\right)^2 +
a_{\alpha\, I}^\dagger\frac{d{a}_\alpha^I}{dt} +
\frac{1}{4} \frac{d\phi_{IJ}}{dt}
\frac{d\phi^{IJ}}{dt}\right].
\label{10}
\eeq
By denoting with $\vec{\pi}$ and $\Pi_{IJ}$ the canonically conjugate momenta
of $\vec{c}$ and $\phi^{IJ}$, respectively, one finds the Hamiltonian
\beq
H=\frac{g^2}{2V} \left[\vec{\pi}^2 +\frac{1}{4}\Pi_{IJ}\Pi^{IJ}\right].
\label{10'}
\eeq
Notice that the fermions do not give any contribution to the energy.
The physical wave functions $\Psi$ are gauge invariant. When reduced on
the toron manifold they become periodic functions of $\vec{c}$, even under
G-parity.
\beq
\Psi(c_i+4\pi/a_i)=\Psi(c_i),\;\;\; \Psi(-\vec{c}\,)=\Psi(\vec{c}\,).
\label{10''}
\eeq
Thus, the spectrum of the momenta $\pi_i$ is quantized, and the
eigenvalues are $a_ik_i/2$, $k_i \in Z$. To find the eigenstates of the
electric flux we recall that the projector over these eigenstates,
${\cal P}[\vec{e}\,]$,
was defined in Eq.~(\ref{8}). It reduces, on the eigenstates of $\pi_i$, to
\beq
{\cal P}[\vec{e}\,]\exp (i k_i c_i a_i/2)=\left\{ \begin{array}{cc}
 \exp (i k_i c_i a_i /2) & {\rm if \;} k_i=e_i\;
{\rm modulo\; 2}, \\ 0 & {\rm otherwise} \end{array} \right.
\label{15}
\eeq

We ought to deal with two problems before being able to write down the
partition function.

The first one is that true physical states are not
eigenstates of $\vec{\pi}$, but rather G-invariant linear combinations of
them. To properly count the multiplicity of the physical states we must
recall that the fermionic wave functions are obtained by applying the
fermion creation operators $a_\alpha^{I\, \dagger}$ to the Fock vacuum
$|0\rangle $. Thus G-parity even fermionic functions contain an even number of
fermions, and G-odd ones an odd number. When $k_i\neq 0$ one finds
physical G-invariant states by combining
the G-odd bosonic wave functions  with all G-odd fermionic wave functions,
and the G-even bosonic wave functions with the G-even fermionic ones.
Fermi statistics allows to find the multiplicity of the fermionic wave
functions of given G-parity. It turns out to be the same for both parities,
namely, 128. When $k_i=0$ only the even bosonic wave functions exist, thus,
the ratio of physical states with $k_i\neq 0$ to states
with $k_i=0$ is 1:2. This result is valid for any component of the momentum
$k_i$ separately.

The second problem involves the zero modes of the scalars $\phi^{IJ}$.
Different values of $\phi^{IJ}$ are not related by 4-D gauge
transformations, thus the spectrum of their momenta is continuous, and their
contribution to the statistical sum~Eq.~(\ref{8}) is divergent.
One may regularize this divergence either by hand, constraining the range of
integration of $\phi^{IJ}$ to a finite volume, or by recalling that N=4
super Yang-Mills comes from the dimentional reduction of N=1 super Yang
Mills in 10 dimensions~\cite{boh}. The scalars are thus the four-dimensional
relics of
ten-dimensional gauge fields. If the compactification radius $R$ of the extra
dimensions is kept finite one finds that the range of integration of the
scalars becomes finite, by the same mechanism which makes the range of
$\vec{c}$ finite. The scalar contribution to the free energy thus is well
defined and finite. In the $R\rightarrow 0$ limit it becomes
\beq
\Tr e^{-\beta H_{\rm scalars}}= \int d\phi_{IJ} d\phi^{IJ} d\Pi_{IJ}
d\Pi^{IJ} e^{-\beta g^2\Pi_{IJ}\Pi^{IJ}/8V}={\rm const}\, R^{-6} \left(
\frac{\beta g^2}{8V}\right)^{-3}.
\label{12}
\eeq
It is crucial to note that since $T_i$ may be written as
$\exp (2\pi i x \tau_3/a_i)$, it acts trivially on $\phi^{IJ}$ (and
$a_\alpha^I$).
Thus the contribution of the scalars to the free energy is an
additive constant {\em independent of the electric flux}; it may be
eliminated by a shift in the definition of the free energy, common to all
flux sectors.

Now, it is immediate to evaluate the statistical sum in Eq.~(\ref{8}) and
find
\beq
\exp\{-\beta F[\vec{e},0]\}= 128 \frac{\beta R^2}{V}
\sum_{\vec{k}\in Z^3}\prod_{i=1}^3
\exp\{(-\pi g^2\beta a_i^2/2V)(k_i +e_i/2)^2\}.
\label{17}
\eeq
The $\vec{e}$-independent normalization factor $ \beta R^2/V $ has been chosen
for convenience. The compactification radius $R$ appears explicitly in this
formula. This is not surprising since $R$ regularizes an infrared divergence.

This formula is already interesting since it means that the free energy of a
nonzero
electric flux behaves as a power. By denoting with $L$ the size of the box
this energy is  $O(g^2/L)$.
This fact suggests that N=4 super Yang-Mills is always in the
Coulomb
phase~\cite{th2}. A splitting $O(g^2/L)$ between the zero-flux sector and
the nonzero-flux ones is
peculiar to supersymmetric theories. In non-supersymmetric models, as pure
Yang-Mills for instance, the classical degeneracy of the toron
configurations~(\ref{11}) is lifted at the quantum level. As a result the
only true vacua are the trivial one $\vec{c}=0$ and its $2^3$
central conjugates~\cite{l}. In this latter case the energy of an electric
flux is zero to all orders in perturbation theory. Indeed, perturbation
theory itself is different, in that the perturbative wave function is localized
around the (discrete) true vacua~\cite{l}.
On the other hand in N=4 supersymmetry
the vacuum degeneracy persists at least to all orders in perturbation
theory.

The free energy $F[0,\vec{m}]$ at $\theta=0$ and $g\gg 1$
(case b) above) can be found as follows.
Strong arguments in favor of the $g\rightarrow 4\pi/g$ duality at
$\theta=0$ were given in ref.~\cite{r}. The core of the reasoning there is
that magnetic monopoles in N=4 can be arbitrarily light: their mass depends
on the scale of gauge symmetry breaking, which is arbitrary, being
associated with a flat direction. The effective Lagrangian of light
composite fields is expected to be renormalizable (this is a consequence of the
so-called ``Veltman's theorem''~\cite{v}).
A N=4 renormalizable Lagrangian is completely
fixed by the gauge group, and the gauge group
that acts on magnetic monopoles is
isomorphic to the original one, which acts on elementary fields.
Moreover, the fact that N=4 supersymmetry gives vanishing beta functions
means that the monopole effective Lagrangian, besides being an infrared
fixed point, it is also an {\em ultraviolet} fixed point. Therefore this
Lagrangian describes correctly monopole dynamics at all scales. This
Lagrangian has the same form of the original one, Eq.~(\ref{2}), with
coupling constant $g_M\equiv 4\pi/g$. We must also remark that,
independently from this argument, the Lagrangian for the
monopole zero
modes (which are the relevant ones for our computation) has been explicitly
found in ref.~\cite{bl}, and it coincides with Eq.~(\ref{10}).
Therefore, one may evaluate $F[0,\vec{m}]$ in complete analogy with the
electric case. The resulting free energy is given by Eq.~(\ref{17}), after
the substitutions $g\rightarrow 4\pi/g$, $\vec{e}\rightarrow \vec{m}$.

Our aim now is to find a S-duality invariant formula for
$F[\vec{e},\vec{m}]$, which reduces to $F[\vec{e},0]$ ($F[0,\vec{m}]$)
in the weak (strong) coupling limit.
The very existence of such an extension is far from obvious; moreover,
$\fre$ must satisfy three non-trivial constraints.
\\
a) It should factorize at $\theta=0$~\cite{th2}: $\fre =F[\vec{e},0] +
F[0,\vec{m}]$.
\\
b) It should account for Witten's phenomenon~\cite{w2}; namely, when
$\theta \rightarrow \theta +2\pi $  the electric flux $\vec{e}$
must transform into
$\vec{e} + \vec{m}$.\\
c) It must obey 't Hooft's duality equations~\cite{th2}. These equations are
due simply to the invariance of the functional integral under discrete
Euclidean O(4) transformation, and read
\beq
\exp\{-\beta F[\tilde{e}, e_3, \tilde{m}, m_3, \tilde{a}, a_3, \beta]\}=
\frac{1}{4} \sum_{\tilde{k},\tilde{l}}
\exp\left\{\pi i (\tilde{l}\cdot \tilde{m}
-\tilde{k}\cdot \tilde{e}) -a_3 F[\tilde{l},e_3,\tilde{k},m_3,
\hat{a},\beta,a_3]\right\}.
\label{19}
\eeq
Here we followed 't Hooft's notations: $\tilde{a}=(a_1,a_2)$, $\hat{a}
=(a_2,a_1)$, $k_i, l_i \in \{ 0, 1\}$.

We are going to present an SL(2, $Z$)-invariant free energy obeying all these
constraints. We believe that the existence of such a free energy is another
strong argument in favor of S-duality. Besides, $\fre$ is a physical
quantity since it is fully gauge-invariant. It is one of the simplest
observables that can be used to investigate a non-Abelian gauge
theory.

Let us define at first the $2 \times 2$ matrix~\cite{s}
\beq
M(\lambda)= \frac{1}{\lambda_2}\left(\begin{array}{cc} 1 & \lambda_1 \\
 \lambda_1 & |\lambda|^2 \end{array} \right), \;\;\; \lambda_1 + i\lambda_2
\equiv iS;\;\; \lambda_1=\frac{\theta}{2\pi},\;\;
\lambda_2=\frac{4\pi}{g^2}.
\label{20}
\eeq
The S-duality in Eq.~(\ref{1}) transforms $M(\lambda)$ as follows
\beq
M\left( \frac{a\lambda + b}{c\lambda + d} \right)=A M(\lambda) A^t,\;\;\;
A=\left( \begin{array}{cc} d & c \\ b & a \end{array} \right ).
\label{21}
\eeq
Thus, an S-duality invariant generalization of Eq.~(\ref{17}) (and
the corresponding one for $\vec{e}=0$, $\vec{m}\neq 0$) is
\bea \exp\{-\beta F[\vec{e},\vec{m},\lambda]\}
&=& 128\frac{\beta R^2}{V}
\prod_{i=1}^3 \sum_{p_i\in Z^2}
\exp \left[ -2\pi\beta \frac{a_i^2}{V}(p_i + f_i)^tM(\lambda)(p_i+f_i)\right].
\label{22}\\
& &  p_i=\left(\begin{array}{c} k_i \\ l_i \end{array} \right);\; k_i,\;
l_i \in Z,\;\;\;
  f_i =\left(\begin{array}{c} e_i/2 \\ m_i/2 \end{array} \right).
\label{22'}
\eea
Eq.~(\ref{22}) is manifestly invariant under the transformation
\beq
\lambda \rightarrow \frac{a\lambda + b}{c\lambda +d} , \;\;\;
\left(\begin{array}{c} e_i/2 \\ m_i/2 \end{array} \right) \rightarrow
\left(\begin{array}{cc} a & -b \\ -c & d \end{array} \right)
\left(\begin{array}{c} e_i/2 \\ m_i/2 \end{array} \right).
\label{22a}
\eeq
Notice that electric and magnetic fluxes transform under duality, as
expected. In particular, under $\lambda \rightarrow 1/\lambda$, $b=c=-1$,
$a=d=0$, and $\vec{e} \leftrightarrow \vec{m}$.

Eq.~(\ref{22}) obeys the constraints a), b) and c).
\\
a) Factorization at $\theta=0$ is manifest.\\
b) Witten's phenomenon is correctly reproduced: under $\lambda \rightarrow
\lambda +1$ the matrix $M(\lambda)$ transforms into
\beq
\left(\begin{array}{cc} 1 & 0 \\ 1 & 1 \end{array} \right) M(\lambda)
\left(\begin{array}{cc} 1 & 1 \\ 0 & 1 \end{array} \right),
\label{22''}
\eeq
and our ansatz Eq.~(\ref{22}) becomes, after a redefinition of the dummy
variable ${p}$,
\bea
\exp\{-\beta F[\vec{e},\vec{m}, \lambda +1 ]\} &=&
\nonumber \\
& = & 128\frac{\beta R^2}{V} \prod_{i=1}^3 \sum_{p_i\in Z^2}
\exp \left[ -2\pi\beta \frac{a_i^2}{V}(p_i + f'_i)^tM(\lambda)(p_i+f'_i)
\right] \nonumber \\ & = &
\exp\{-\beta F[\vec{e}+ \vec{m},\vec{m}, \lambda]\},
\label{23}
\eea
with ${f'}_i^t=(e_i/2 + m_i/2, m_i/2)$.
\\
c) The least evident property that must be obeyed by
$F[\vec{e},\vec{m},\lambda]$ is 't Hooft duality. To prove Eq.~(\ref{19})
one may take, for simplicity, $e_3, m_3 =0$. The right hand side of
Eq.~(\ref{19}) then becomes
\bea
& & 128\frac{1}{4} \sum_{q_i}\sum_{k\in Z^6}
\exp \left\{2\pi i (q_1^t B f_1 + q_2^t B f_2)
-2\pi \frac{a_3 a_2}{\beta a_1} (k_1 +
q_1/2)^t M(\lambda) (k_1 + q_1/2) \right. \nonumber \\ &  &
\left.
-2\pi \frac{a_3 a_1}{\beta a_2} (k_2 + q_2/2)^t M(\lambda) (k_2 + q_2/2)
-2\pi \frac{a_3 \beta}{a_1 a_2} (k_3 + f_3)^t M(\lambda) (k_3 + f_3)
\right\}. \nonumber \\ & &
\label{24}
\eea
The notations here are as follows
\beq
B=\left( \begin{array}{cc} 0 & 1 \\ -1 & 0 \end{array} \right),\;\;\;
q_i=\left( \begin{array}{c} r_i \\ s_i  \end{array} \right), \;\;\; r_i,\;
s_i \in \{ 0,1\}.
\label{25}
\eeq
Let us perform a Poisson resummation on the directions $i=1,2$
\bea
& &\sum_{q_i}\sum_{k\in Z^4}
\exp\left\{
2\pi i (q_1^t B f_1 + q_2^t B f_2) -2\pi \frac{a_3 a_2}{\beta a_1} (k_1 +
q_1/2)^t M(\lambda) (k_1 + q_1/2) \right. \nonumber \\ & & \left.
-2\pi \frac{a_3 a_1}{\beta a_2} (k_2 + q_2/2)^t M(\lambda) (k_2 + q_2/2)
\right\} =
\frac{\beta^2}{4a_3^2} \sum_{q_i} \sum_{k\in Z^4}
\exp\left\{2\pi i(q_1^t B f_1 +
\right. \nonumber \\ & & \left.
 + q_2^t B f_2 +    k_1^t q_1/2 + k_2^t q_2/2)
-\frac{\pi}{2} \frac{\beta a_1}{a_2 a_3} k_1^t M(\lambda)^{-1} k_1
-\frac{\pi}{2} \frac{\beta a_2}{a_1 a_3} k_2^t M(\lambda)^{-1} k_2\right\}.
\label{26}
\eea
Using the fact that $M^{-1}(\lambda)= B M(\lambda) B^t$, and that
\beq
\frac{1}{16}\sum_{q_i} \exp[2\pi i(q_1^t B f_1 + q_2^t B f_2 + k_1^t q_1/2
+ k_2^t q_2/2)] =
\left\{ \begin{array}{cc} 1 & {\rm if}\; k_i= -2 B f_i \;{\rm
mod\; 2} \\ 0 & {\rm otherwise} \end{array} \right.,
\label{27}
\eeq
one finds that, by substituting Eq.~(\ref{26}) into Eq.~(\ref{24}),
this latter expression equals the left hand side of 't Hooft duality
equation~(\ref{19}).

This completes the demostration that our ansatz for the free energy does
satisfy the three relevant physical constraints mentioned above.

The generalization of our ansatz Eq.~(\ref{22}) to any simply-laced Lie
group is straightforward, as well as the verification of 't Hooft's duality
equations; these topics shall be dealt with in a forthcoming
publication~\cite{ggpz}. As previously remarked, the main difficulty in
extending our analysis from SU(2) to an arbitrary simply-laced Lie group is
to count properly the physical states. These states are invariant under the
Weyl group of the Lie algebra; the correct multiplicities of those states,
for a N=4 theory, are not as obvious as in our SU(2) example.

We would like to conclude by stressing again that in  our ansatz the N=4
supersymmetric gauge theory lies in a Coulomb phase, without mass gap.
In this phase the theory realizes the symmetry in Eq.~(\ref{1}) in
a self-dual way. For instance, the theory can be smoothly deformed from a
weak-coupling region ($g\ll 1$, $\theta=0$) to a strong-coupling one
($g\gg 1$, $\theta=0$) without undergoing a phase transition. This property
is peculiar to the Coulomb phase, as remarked firstly by 't
Hooft~\cite{th2}.
\vskip .2in
\noindent
Acknowledgements
\vskip .1in
\noindent
We would like to thank M. Ro\v{c}ek for useful comments on the manuscript;
one of us (L.G.) would like to thank the Department of Physics of NYU for its
kind hospitality. This work is supported in part by MURST and INFN, Italy. A.G.
is supported in part by the BSF (American-Israeli Bi-National Science
Foundation) and an Alon Fellowship;
M.P. is supported in part by NSF under grant no. PHY-9318781.

\end{document}